\def\rfr#1{eq. (\ref{#1})}

\def\asec{$''$ cy$^{-1}$}

\def\dert#1#2{\frac{{{d}}{#1}}{{{d}}{#2}}}              

\def\asec{$''$ cy$^{-1}$}

\def\bar{\begin{eqnarray}}
\def\ear{\end{eqnarray}}
\def\bb{\bibitem}
\def\eqi{\begin{equation}}
\def\eqf{\end{equation}}
\def\eqia{\begin{eqnarray}}
\def\eqfa{\end{eqnarray}}
\def\rp#1#2{{#1\over#2}}

\def\lb#1{\label{#1}}

\def\oc2{$\mathcal{O}(c^{-2})$}

\documentclass{aastex}
\usepackage{url}

\RequirePackage{color}

\begin{document}

\title{Putting Yukawa-like Modified Gravity (MOG) on the test in the Solar System}

\shorttitle{Putting Yukawa-like Modified Gravity (MOG) on the test in the Solar System}
\shortauthors{L. Iorio}

\author{Lorenzo Iorio }
\affil{INFN-Sezione di Pisa. Permanent address for correspondence: Viale Unit\`{a} di Italia 68, 70125, Bari (BA), Italy. E-mail: lorenzo.iorio@libero.it}

\begin{abstract}
We deal with a Yukawa-like long-range modified model of gravity (MOG) which recently allowed  to successfully accommodate many astrophysical and cosmological features without resorting to dark matter. On Solar System scales MOG predicts retrograde secular precessions of the planetary longitudes of the perihelia $\varpi$ whose existence has been put on the test here by taking the ratios of the observationally estimated Pitjeva's corrections to the standard Newtonian/Einsteinian perihelion precessions for different pairs of planets. It turns out that MOG, in the present form which turned out to be  phenomenologically successful on astrophysical scales, is ruled out at more than $3\sigma$ level in the Solar System. If and when other teams of astronomers will independently estimate their own corrections to the usual precessions of the perihelia it will be possible to repeat such a test.
 \end{abstract}


\section{Introduction}\lb{intro}
The modified gravity (MOG) theory put forth in \citep{Mof06}  was used successfully to describe various observational phenomena on astrophysical and cosmological scales without resorting to dark matter (see \cite{Mof08} and references therein). It is a fully covariant theory of gravity which is based on the existence of a massive vector field  coupled universally to matter. The theory yields a Yukawa-like modification of gravity with three  constants which, in the most general case, are running;
they are present in the theory's action as scalar fields which represent the gravitational constant, the vector field coupling constant and the vector field mass. Actually, the issue of the running of the parameters of modified models of gravity is an old one, known in somewhat similar  contexts since the early 1990s (see, e.g., \citep{Ber93} and references therein).
An approximate solution of the MOG field equations \citep{Mof07} allows to compute their values  as functions of the source's mass.

The resulting Yukawa-type modification of the inverse-square Newton's law in the gravitational field of a central mass $M$ is \citep{Mof07,Mof08}
\eqi A_{\rm MOG} = -\rp{G_{\rm N}M}{r^2}\left\{1 + \alpha\left[1-(1+\mu r)\exp(-\mu r)\right]\right\}, \lb{MOG_acc}\eqf

where    $G_{\rm N}$ is the Newtonian gravitational constant and \citep{Mof07,Mof08}

\eqi
\alpha = \rp{M}{\left(\sqrt{M} + C_1^{'}\right)^2}\left(\rp{G_{\infty}}{G_{\rm N}}-1\right),\ G_{\infty}\approx 20\ G_{\rm N},\ C_1^{'}\approx 25000\ {\rm M}^{1/2}_{\odot},\lb{MOG_alfa}
\eqf

\eqi
\mu = \rp{C_2^{'}}{\sqrt{M}},\ C_2^{'}\approx 6250\ {\rm M}^{1/2}_{\odot}\ {\rm kpc}^{-1}.\lb{MOG_mu}
\eqf
Such values have been obtained by \citep{Mof07} as a result of the fit of the velocity rotation curves of some galaxies in the framework of the searches for an explanation of the flat rotation curves of galaxies without resorting to dark matter.

In this paper we will put \rfr{MOG_acc} on the test in the Solar System in order to check if it is  compatible with the latest  determinations of the corrections $\left<\Delta\dot\varpi\right>$ to the usual Newtonian/Einsteinian planetary perihelion precessions \citep{Pit05a,Pit05b}.  Such quantities are solved-for parameters phenomenologically estimated in a least-square sense by fitting about one century of observations of various kinds concerning planetary motions  with their values theoretically predicted by simultaneously integrating the equations of motions of the major bodies of the Solar System written in terms of a complete suite of dynamical force models. They encompass all the known Newtonian effects along with those predicted by general relativity, except for the Lense-Thirring effect \citep{LT} which was left unmodelled (see below). Thus, $\left<\Delta\dot\varpi\right>$  account for any unmodelled/mismodelled dynamical features of motion  and are suitable, in principle, to put on the test MOG, provided that they are used in a suitable way.

Note that \citet{Mof08} explicitly write that \rfr{MOG_acc}, with \rfr{MOG_alfa} and \rfr{MOG_mu}, is not in contradiction with the present-day knowledge of Solar System dynamics. We will show that it is not so also for any other (non-zero) values of $\alpha$ and $\mu$, with the only quite general condition that $\mu r\ll 1$ in Solar System, as it must be for any long-range modified model of gravity. It is interesting to point out that Yukawa-like modifications of Newton's law might also be tested in the context of stellar dynamics \citep{Ber05}.
Here we outline the procedure that we will follow.

Generally speaking,
let ${\rm LRMOG}$ (Long-Range Modified Model of Gravity) be a given exotic model of modified gravity parameterized in terms of, say, $K$, in a such a way that $K=0$ would imply no modifications of gravity at all. Let ${\mathcal{P}}({\rm LRMOG})$ be the prediction of a certain effect induced by such a model like, e.g., the secular precession of the perihelion of a planet. For all the exotic models considered it turns out that\footnote{In our case it will be $K=-\alpha\mu^2$, as we will see in Section \ref{seconda}.} \eqi{\mathcal{P}}({\rm LRMOG}) = Kg(a,e),\eqf where $g$ is a function of the system's orbital parameters $a$ (semimajor axis) and $e$ (eccentricity); such $g$ is a peculiar consequence of the model ${\rm LRMOG}$ (and of all other models of its class with the same spatial variability).
Now, let us take the ratio of ${\mathcal{P}}({\rm LRMOG})$ for two different systems {\rm A} and {\rm B}, e.g. two Solar System's planets: ${\mathcal{P}}_{\rm A}({\rm LRMOG})/{\mathcal{P}}_{\rm B}({\rm LRMOG}) = g_{\rm A}/g_{\rm B}$. The model's parameter $K$ has now been canceled, but we still have a prediction that retains a peculiar signature of that model, i.e. $g_{\rm A}/g_{\rm B}$. Of course, such a prediction is valid if we assume $K$ is not zero, which is just the case both theoretically (${\rm LRMOG}$ is such that should $K$ be zero, no modifications of gravity at all occurred) and observationally because $K$ is usually determined by other independent long-range astrophysical/cosmological observations. Otherwise, one would have the meaningless prediction $0/0$. The case $K=0$ (or $K\leq\overline{K}$) can be, instead, usually tested  by taking one perihelion precession at a time.
If we have observational determinations ${\mathcal{O}}$ for {\rm A} and {\rm B} of the effect considered above  such that they are affected  also\footnote{If they are differential quantities constructed by contrasting observations to predictions obtained by analytical force models of canonical Newtonian/Einsteinian effects, ${\mathcal{O}}$ are, in principle, affected also by the mismodelling in them.} by  ${\rm LRMOG}$ (it is just the case for the purely phenomenologically estimated corrections to the standard Newton-Einstein perihelion precessions, since ${\rm LRMOG}$ has not been included in the  dynamical force models of the ephemerides adjusted to the planetary data in the least-square parameters' estimation process by Pitjeva \citep{Pit05a,Pit05b}), we can construct ${\mathcal{O}}_{\rm A}/\mathcal{O}_{\rm B}$ and compare it with the prediction for it by ${\rm LRMOG}$, i.e. with $g_{\rm A}/g_{\rm B}$. Note that $\delta{\mathcal{O}}/{\mathcal{O}}>1$ only means that ${\mathcal{O}}$ is compatible with zero, being possible a nonzero value smaller than $\delta{\mathcal{O}}$. Thus, it is perfectly meaningful to construct ${\mathcal{O}}_{\rm A}/\mathcal{O}_{\rm B}$. Its uncertainty will be conservatively evaluated as $|1/{\mathcal{O}}_{\rm B}|\delta{\mathcal{O}}_{\rm A} + |{\mathcal{O}}_{\rm A}/{\mathcal{O}}_{\rm B}^2|\delta{\mathcal{O}}_{\rm B}$. As a result, ${\mathcal{O}}_{\rm A}/\mathcal{O}_{\rm B}$ will be compatible with zero. Now, the question is: Is it the same for $g_{\rm A}/g_{\rm B}$ as well? If yes, i.e. if \eqi\rp{{\mathcal{O}}_{\rm A}}{\mathcal{O}_{\rm B}}=\rp{{\mathcal{P}}_{\rm A}({\rm LRMOG})}{{\mathcal{P}}_{\rm B}({\rm LRMOG})}\eqf within the errors, or, equivalently, if \eqi\left|\rp{{\mathcal{O}}_{\rm A}}{\mathcal{O}_{\rm B}} - \rp{{\mathcal{P}}_{\rm A}({\rm LRMOG})}{{\mathcal{P}}_{\rm B}({\rm LRMOG})}\right|=0\eqf within the errors,  ${\rm LRMOG}$ survives (and the use of the single perihelion precessions can be used to put upper bounds on $K$). Otherwise, ${\rm LRMOG}$ is ruled out.

\section{The predicted perihelion precessions and the confrontation with the measured non-standard rates}\lb{seconda}
In the case of the Sun, \rfr{MOG_alfa} and \rfr{MOG_mu} yield
\eqi\alpha_{\odot}\approx 3\times 10^{-8},\ \mu\approx 3\times 10^{-5}\ {\rm  AU}^{-1},\eqf
so that \eqi\alpha_{\odot}\mu^2 = 3\times 10^{-17}\ {\rm  AU}^{-2}.\lb{bastaa}\eqf
Since in the Solar System $\mu r \approx 10^{-5}-10^{-4},$ we can safely assume
$ \exp(-\mu r)\approx 1-\mu r,$
so that
\eqi A_{\rm MOG}\approx  -\rp{G_{\rm N}M}{r^2}\left(1+\alpha\mu^2 r^2\right).\lb{accpicc}\eqf As a result, a radial, uniform perturbing acceleration
\eqi A = -G_{\rm N}M\alpha\mu^2\approx 10^{-19}\ {\rm m}\ {\rm s}^{-2}\lb{piccola}\eqf is induced.

The secular, i.e. averaged over one orbital revolution,
effect of a small radial and unform perturbing acceleration on the longitude of the perihelion of
a planet $\varpi$ has been worked out by, e.g., \citet{San06}; it amounts to
\eqi\left<\dert\varpi t\right> = A\sqrt{\rp{a(1-e^2)}{G_{\rm N}M}}=-\alpha\mu^2\sqrt{ G_{\rm N}M a (1-e^2)}.\lb{prece}\eqf
Clearly, using only one perihelion rate at a time would yield no useful information on MOG due to the extreme smallness of the perturbing acceleration, as told us by \rfr{piccola}. Thus, let us take the ratios of the perihelion precessions. It must be noted that the following analysis is, in fact, truly independent of the values of $\alpha$ and $\mu$, provided only that $\alpha\mu^2 r^2\ll 1$ in the Solar System so as that the perturbative approach can be applied to \rfr{accpicc}; the condition $\mu r\ll 1$ is the cornerstone of any long-range modified models of gravity, and should $\alpha\approx 1$ the planetary orbits would have been distorted in a so huge manner that it would have been detected since long time.
Applying the scheme outlined in Section \ref{intro} to our case in which $K=-\alpha\mu^2$ and $g(a,e)=\sqrt{G_{\rm N}M a(1-e^2)}$, one can construct
\eqi \Pi\equiv\rp{\left<\Delta\dot\varpi_{\rm A}\right>}{\left<\Delta\dot\varpi_{\rm B}\right>}\eqf with the estimated corrections $\left<\Delta\dot\varpi\right >$ to the standard Newtonian/Einsteinian perihelion precessions of planets A and B, listed in Table \ref{tavola1},   and compare them to  the theoretical prediction
\eqi {\mathcal{A}}\equiv
\sqrt{   \rp{a_{\rm A}(1-e^2_{\rm A})} {a_{\rm B}(1-e^2_{\rm B})} },\eqf obtained from \rfr{prece}, for that pair of planets A and B.
\begin{table}[!h]
\caption{ Inner planets. First row: estimated corrections to the standard precessions of the longitudes of the perihelia in $10^{-4}$ \asec\ (\asec$\rightarrow$ arcseconds per century), from Table 3 of \citep{Pit05b} (apart from Venus). The quoted
errors, in $10^{-4}$ \asec, are not the formal ones but are  realistic. The formal errors  are quoted in square brackets (E.V. Pitjeva, personal communication to the author, November 2005). The units are
$10^{-4}$ \asec. Second row: semimajor axes, in Astronomical Units (AU). Their formal errors are in Table IV of \citep{Pit05a}, in m. Third row: eccentricities. Fourth row: orbital periods in years. The result for Venus have been recently obtained by including the Magellan radiometric data (E.V. Pitjeva, personal communication to the author, June 2008).\label{tavola1} }

\centering
\bigskip
\begin{tabular}{ccccc}
\hline\noalign{\smallskip}
& Mercury & Venus & Earth & Mars\\
\noalign{\smallskip}\hline\noalign{\smallskip}
$\left<\Delta\dot\varpi\right>$ ($10^{-4}$\ \asec) & $-36\pm 50[42]$ & $-4\pm 5[1]$& $-2\pm 4[1]$ & $1\pm 5[1]$\\
$a$ (AU) & 0.387 & 0.723 & 1.000 & 1.523 \\
$e$ & 0.2056 & 0.0067 & 0.0167  & 0.0934\\
$P$ (yr) & 0.24 & 0.61 &  1.00 & 1.88\\
\noalign{\smallskip}\hline
\end{tabular}

\end{table}
%
%
%
%
%
%
%
%
%
%
The results are in Table \ref{tavola2}.
\begin{table}[!h]
\caption{ First column: pair of planets. Second column: $\Pi$ for that pair of planets. The errors come from the realistic uncertainties in $\left<\Delta\dot\varpi\right>$. Third column: $\mathcal{A}$ for that pair of planets. Fourth column: $\sigma$ level of discrepancy between $\Pi$ and
$\mathcal{A}$  for that pair of planets.\label{tavola2}}
\centering
\bigskip
\begin{tabular}{cccc}
\hline\noalign{\smallskip}
A B & $\Pi$ & $\mathcal{A}$ & $\sigma$\\
\noalign{\smallskip}\hline\noalign{\smallskip}

Venus Mercury & $0.1\pm 0.3$ & 1.4 & 4\\
Earth Mercury & $0.05 \pm 0.18$ & 1.64 & 8\\
Mars Mercury & $-0.03\pm 0.18$ & 2.02 & 11\\
 %
\noalign{\smallskip}\hline
\end{tabular}
\end{table}
$\left|\Pi - \mathcal{A}\right|$ is different from zero at more than $3\sigma$ level for A = Venus, B = Mercury, A = Earth, B = Mercury and A = Mars, B = Mercury. It is important to note that the errors have been conservatively evaluated as
\eqi\delta\Pi\leq \left|\Pi\right|\left(\rp{\delta\left<\Delta\dot\varpi_{\rm A}\right>}{|\left<\Delta\dot\varpi_{\rm A}\right>|}
+\rp{\delta\left<\Delta\dot\varpi_{\rm B}\right>}{|\left<\Delta\dot\varpi_{\rm B}\right>|}
\right)\eqf  because of the existing correlations\footnote{The maximum correlation, $26\%$, occurs for the Earth and Mercury (E.V. Pitjeva, personal communication to the author, November 2005).} among the estimated corrections to the precessions of perihelia.

If we repeat our analysis by subtracting from $\left<\Delta\dot\varpi\right>$ the main canonical unmodelled effect, i.e. the general relativistic Lense-Thirring precessions \citep{LT} induced by the Sun's angular momentum \citep{Ior07} shown in Table \ref{scassa}, i.e. if we use
\eqi\Pi^{\star}\equiv\rp{\left<\Delta\dot\varpi_{\rm A}\right>^{\star} }{\left<\Delta\dot\varpi_{\rm B}\right>^{\star}}= \rp{ \left<\Delta\dot\varpi_{\rm A}\right> - \dot\varpi^{(\rm LT)}_{\rm A}}{\left<\Delta\dot\varpi_{\rm B}\right> - \dot\varpi^{(\rm LT)}_{\rm B}},\eqf
the situation does not substantially change, apart from the sigma level at which $\left|\Pi^{\star}-\mathcal{A}\right|$  is not compatible with zero, as shown in Table \ref{tavola3}.
\begin{table}[!h]
\caption{ General relativistic Lense-Thirring precessions of the longitudes of perihelia of the inner planets of the Solar System in $10^{-4}$ \asec.\label{scassa}
}
\centering
\bigskip
\begin{tabular}{ccccc}
\hline\noalign{\smallskip}
& Mercury & Venus & Earth & Mars\\
\noalign{\smallskip}\hline\noalign{\smallskip}
 $\dot\varpi^{(\rm LT)}$ ($10^{-4}$ \asec)& $-20$ & $-3$ & $-1$ & $-0.3$\\ 
\noalign{\smallskip}\hline
\end{tabular}
\end{table}
\begin{table}[!h]
\caption{ First column: pair of planets. Second column: $\Pi^{\star}$ for that pair of planets including the unmodelled general relativistic Lense-Thirring effect. The errors come from the realistic uncertainties in $\left<\Delta\dot\varpi\right>$. Third column: $\mathcal{A}$ for that pair of planets. Fourth column: $\sigma$ level of discrepancy between $\Pi$ and
$\mathcal{A}$  for that pair of planets.\label{tavola3}}
\centering
\bigskip
\begin{tabular}{cccc}
\hline\noalign{\smallskip}
A B & $\Pi^{\star}$ & $\mathcal{A}$ & $\sigma$\\
\noalign{\smallskip}\hline\noalign{\smallskip}
Venus Mercury & $0.06\pm 0.51$ & 1.4 & 2.7\\
Earth Mercury & $0.06 \pm 0.44$ & 1.64 & 3.5\\
Mars Mercury & $-0.08\pm 0.56$ & 2.02 & 3.7\\
 %
\noalign{\smallskip}\hline
\end{tabular}
\end{table}
%

The availability of the corrections to the usual rates of perihelia of several planets allows us to put on the test MOG also in another way as well.
The acceleration law of \rfr{MOG_acc} can also be recast in the commonly used Yukawa form \citep{Mof07}
\eqi A_{\rm Y} = -\rp{G_{\rm Y}M}{r^2}\left[1+\alpha_{\rm Y}\left(1+\rp{r}{\lambda}\right)\exp\left(-\rp{r}{\lambda}\right)\right],\lb{YUK_acc}\eqf
where
\eqi G_{\rm Y}=\rp{G_{\rm N}}{1+\alpha_{\rm Y}},\eqf
\eqi \alpha_{\rm Y}=-\rp{(G_{\infty}-G_{\rm N})M}{(G_{\infty}-G_{\rm N})M + G_{\rm N}(\sqrt{M}+C_1^{'})^2},\eqf
\eqi\lambda = \rp{1}{\mu}.\eqf
In the case of the Sun
\eqi\alpha^{\odot}_{\rm Y}=-3.04\times 10^{-8}, \ G_{\rm Y} = 1.00000003040 G_{\rm N}, \ \lambda = 33000\ {\rm AU}.\lb{YUK_param}\eqf

A Yukawa-type acceleration of the form of \rfr{YUK_acc} has been tested by \citet{Ior07b} in the Solar System without any a-priori assumption on the size of\footnote{The strength parameter $\alpha$ used in \citep{Ior07b} can be identified with $\alpha_{\rm Y}$ here.} $\alpha_{\rm Y}$; concerning $\lambda,$ it was only assumed that $\lambda\gtrsim ae$. By using the corrections to the standard rates of the perihelia of A = Earth and B =  Mercury quoted in Table \ref{tavola1} \citet{Ior07b} found \eqi\lambda = \rp{a_{\rm B} - a_{\rm A}}{\ln\left(\sqrt{\rp{a_{\rm B}}{a_{\rm A}}}\Pi\right)}= 0.182\pm 0.183\ {\rm AU},\lb{YUK_lam}\eqf which contradicts \rfr{YUK_param}.
Using the data for Venus in  the equation for $\alpha_{\rm Y} $\citep{Ior07b} \eqi\alpha_{\rm Y} = \rp{2\lambda^2\left<\Delta\dot\varpi\right>}{\sqrt{G_{\rm Y}Ma}}\exp\left(\rp{a}{\lambda}\right)\lb{YUK_al}\eqf
yields\footnote{According to \rfr{YUK_param}, using $G_{\rm N}$ in \rfr{YUK_al} instead of $G_{\rm Y}$, as done in \citep{Ior07b}, does not produce appreciable modifications of the results.}
\eqi \alpha_{\rm Y} = (-1\pm 4 )\times 10^{-11},\eqf which is three orders of magnitude smaller than the result of \rfr{YUK_param}.

If we use $\Pi^{\star}$ for the Earth and Mercury in \rfr{YUK_lam} and $\left<\Delta\dot\varpi\right>^{\star}$ for Venus in \rfr{YUK_al} the results does not change appreciably; indeed, we have
\eqi\lambda = 0.2\pm 0.4\ {\rm AU},\ \alpha_{\rm Y}= (-0.3 \pm 2.7)\times 10^{-11}.\eqf

\section{Conclusions}
In the framework of the attempts of explaining certain astrophysical and cosmological features without invoking dark matter, MOG \citep{Mof06,Mof07} is a long-range modified model of gravity, based on a vector field and three scalar fields representing running constants, which assumes a Yukawa-like form. Recent developments of this theory allowed their proponents to fix \citep{Mof07,Mof08} the values of the constants entering it. We have shown that, on Solar System scales, MOG yields a uniform anomalous acceleration which would induce retrograde planetary perihelion precessions. We put on the test the possibility that such exotic precessions exist by comparing the ratio of them $\mathcal{A}$ for different pairs of planets to the ratio $\Pi$ of the corrections to the usual Newtonian/Einsteinian precessions estimated by E.V. Pitjeva which account for any unmodelled/mismodeleld dynamical effects. It turns out that $\Pi \neq \mathcal{A}$ at more than $3\sigma$ level even by including in $\Pi$ the main unmodelled canonical effect, i.e. the general relativistic Lense-Thirring precessions. Conversely, using the estimated corrections to the planetary perihelion rates to phenomenologically determine the strength parameter of the putative MOG Yukawa force and its range yields values which are neatly incompatible with those of MOG \citep{Mof07,Mof08}. In assessing the results presented here it must be considered that, at present, no other people have estimated the non-standard part of the planetary perihelion motions; it would certainly be useful to repeat the present analysis if and when other teams of astronomers will estimate their own set of corrections to the standard perihelion precessions as well.


\end{document}